\documentclass[useAMS,usenatbib]{mn2e}
\usepackage{graphicx}
\usepackage{epsfig}

\begin{document}
\title{Calibrated Estimates of the Energy in Major Flares of GRS 1915+105}
\author[Brian Punsly and J\'{e}r$\hat{\mathrm{o}}$me Rodriguez] {Brian Punsly and J\'{e}r$\hat{\mathrm{o}}$me Rodriguez\\ 1415 Granvia Altamira, Palos Verdes Estates CA, USA
90274 and ICRANet, Piazza della Repubblica 10 Pescara 65100, Italy\\
\\Laboratoire AIM, CEA/DSM-CNRS-Universit\'{e} Paris Diderot, IRFU
SAp, F-91191 Gif-sur-Yvette, France\\
 \\E-mail: brian.punsly1@verizon.net or brian.punsly@comdev-usa.com}

\maketitle \label{firstpage}
\begin{abstract} We analyze the energetics of the major radio flare
of October 8 2005 in GRS 1915+105. The flare is of particular
interest because it is one of the most luminous and energetic radio
flares from a Galactic black hole that has ever been observed. The
motivation is two-fold. One, to learn more about the energetics of
this most extreme phenomenon and its relationship to the accretion
state. The second is to verify if the calibrated estimates of the
energy of major radio flares (based on the peak low frequency
optically thin flux) derived from flares in the period 1996-2001 in
Punsly \& Rodriguez (2013), PR13 hereafter, can be used to estimate
plasmoid energy beyond this time period. We find evidence that the
calibrated curves are still accurate for this strong flare.
Furthermore, the physically important findings of PR13 are supported
by the inclusion of this flare: the flare energy is correlated with
both the intrinsic bolometric X-ray luminosity, $L_{\mathrm{bol}}$,
$\sim 1$ hour before ejection and $L_{\mathrm{bol}}$ averaged over
the duration of the ejection of the plasmoid and $L_{\mathrm{bol}}$
is highly elevated relative to historic levels just before and
during the ejection episode. A search of the data archives reveal
that only the October 8 2005 flare and those in PR13 have adequate
data sampling to allow estimates of both the energy of the flare and
the X-ray luminosity before and during flare launch.
\end{abstract}
\section{Introduction}
Accretion and the  associated ejection processes are ubiquitous
phenomena of our Universe. They are indeed seen in many
astrophysical objects: they power the distant gamma-ray bursts, the
massive black holes lurking at the center of active galaxies. Closer
to us accretion/ejection is also a source of radiations in young
stellar objects and in Galactic accreting compact objects also
referred to as 'microquasars'. Understanding the physics of
accretion and jets and also their (potential) links is thus of
primal importance to understand a large range of celestial objects.
In this respect microquasars present several advantages over the
other aforementioned sources. They are close and (very) bright  in
most wavelengths which makes them easy to observe and follow, and
they also vary on short human-followable time scales (from ms to
year).
\par The black hole GRS~1915+105 launches more
superluminal radio flares out to large distances than any other
Galactic object \citep{mir94,fen99,dha00}. The incredibly large
energy of the major ejections pushes our understanding of the
physics of jet launching in black hole accretion systems to the
limit. Great progress was made towards establishing phenomenological
relationships between the accretion flow before and during flare
launch and the power of the relativistic ejections in Punsly \&
Rodriguez (20131a, PR13 hereafter). However, our understanding is
far from complete. The most energetic radio flares are the most
enigmatic from a physical perspective. It is very unclear how an
accretion flow around a black hole can eject so much energy
\citep{pun14}. Very strong radio flares ($\sim$ the Eddington
luminosity) are rare and few of these have information regarding the
accretion flow from serendipitous X-ray observations just before and
during the ejection process.
\par Microquasars exhibit two distinct types of
jets: discrete ejections of 'plasmoids' \citet{mir94}, the major
flares that are the subject of this study, and the so-called compact
jet that is present during the X-ray hard state
\citep{cor00,cor03,gal03}. There is no established relationships
between these two outflow states and the physical connection between
the two phenomena, if any, is not well understood. Huge
multi-wavelength efforts have been made in the past twenty years to
try and understand the origin of all type of jets and their
connection to the accretion processes (e.g.
\citet{mir98,cor00,cor03,kle02,gal03,rod08,rod09,rus11}). Still, the
physical relationship between the accretion flow and the production
of jetted outflows remans speculative. This is especially so for the
powerful major ejections. In PR13, it was pointed out that X-ray
observations with time resolution on the order of days are too
coarsely spaced to resolve the X-ray state before and during the
brief major flare ejection episodes that occur unexpectedly a few
times a year. As a consequence, any X-ray data that is coincident
with the instant of major ejection launching is purely
serendipitous. Culling through large RXTE data sets, it was
demonstrated in PR13 that empirical relationships exist between the
accretion states and major flare ejections. In particular, the X-ray
luminosity is highly elevated in the last hours preceding major
ejections and it is correlated with the power required to eject the
discrete plasmoids. Secondly, the X-ray luminosity was found to be
highly variable during the ejection of the plasmoids, but the time
averaged X-ray luminosity during the ejection event is correlated
with that just before the plasmoid is launched and is of a similar
(but perhaps a slightly lower) level. Thusly motivated, we seek to
expand the database of PR13 that showed the correlated disk-jet
behavior.
\par In this paper, we study one such extremely powerful major radio
flare, that of October 8 2005 (MLD 53651). We estimate that is the
fifth or sixth strongest radio flare ever detected in the 20 years
of monitoring GRS 1915+105. Of these 6 radio flares only this flare
and the radio flare launched on April 13 1998 (MJD 50916) have X-ray
observations just before and during the launching of the ejection.
This strong radio flare provides an excellent test case to validate
the empirical relationships between the accretion state before and
during radio flare launch and the energy of the major radio flare
that were found in PR13. The paper is organized as follows. Section
2 describes the estimation of the radio flare ejection time. Section
3 is an estimate of the intrinsic X-ray luminosity from 1.2 keV - 50
keV, $L_{\mathrm{bol}}$, before and during the launch of the
plasmoid. The following section is a computation of the energy of
the plasmoid, $E$, associated with the radio flare.  Section 5 will
compare the results of Sections 3 and 4 to the empirical
relationships in PR13.
\section{The Time of Ejection Launch}
Determining the time of the ejection is essential. It allows us to
establish a temporal (and perhaps causal) chain of events and the
fluence. This time signature provides a physical context for the
individual X-ray observations. Every optically thin radio flare is
preceded by a rise in optically thick high frequency radio emission.
As the ejected plasmoid expands, the optical depth to synchrotron
self absorption (SSA) decreases and the spectrum steepens at ever
decreasing frequency until it is optically thin at low frequency. In
PR13 it was shown that the 15 GHz light curve (from the Ryle
Telescope public archive http://www.mrao.cam.ac.uk/~guy/1915/) can
provide an excellent estimate of the time that the plasmoids were
ejected. This is true provided that the linear extrapolation of the
light curve backwards in time to the background flux density level
is sufficiently short. This was verified both empirically by the
agreement of this technique with plasmoid ejections times deduced
from radio interferometry data and also with theoretical arguments
in PR13. One reason for choosing the radio flare on MJD 53651 (MJD
will be dropped hereafter) is the excellent launch time estimate,
the details of which are illustrated in Figure 1. We have no
estimate of the background flux density level because GRS 1915+105
was very active preceding this radio flare. However, the rise is
very steep and to a very high level. Thus, inspection of Figure 1
indicates that the only plausible extrapolations to a nonzero
background flux level yields a start time between 53650.78 and
5360.83 (a mere 1 hour uncertainty).

\begin{figure}
\includegraphics[width= 85 mm, angle= 0]{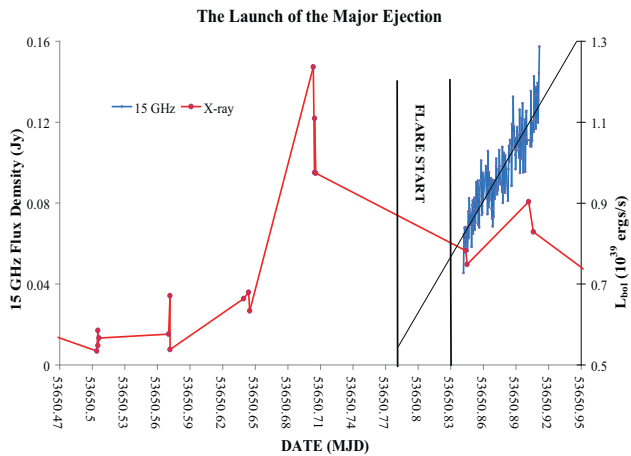}
\caption{A detailed view of the 15 GHz flux density and the
intrinsic X-ray luminosity, $L_{\mathrm{bol}}$, light curves near
the estimated ejection time for the MJD 53651 radio flare.
$L_{\mathrm{bol}}$ is estimated from the RXTE ASM data per the
methods of PR13 as detailed in Section 3.}
\end{figure}
\section{The Intrinsic X-ray Luminosity}
An important result from PR13 was the development and validation of
a method to estimate the intrinsic X-ray luminosity from 1.2 keV -
50 keV, $L_{\mathrm{bol}}$, from the ASM data of RXTE. The estimates
of $L_{\mathrm{bol}}$ are based on models where the main
contribution is due to thermal Comptonisation of soft (cold $\sim
0.2$ keV) photons by hot ($\sim$ 20-100 keV) electrons present in a
so-called corona. The important parameters to estimate
$L_{\mathrm{bol}}$ are $\mathrm{kT_{inj}}$, $\mathrm{kT_{e}}$,
$\tau$ and the Comptonised normalization (see Section 4.2.2 of
PR13). The method was verified by finding PCA observations that are
modeled as such and comparing them to quasi-simultaneous ASM data.
The estimator was derived using $\chi$ class spectra as defined by
\citet{bel00}, but empirically the estimator applies fairly
accurately to all the PCA models of states that it was tested
against except the very soft states. Comparison to the PCA generated
models indicate a very small stochastic relative error of 13\% -
14\%. The main systematic error in $L_{\mathrm{bol}}$ is in the
column density, $N_{H}$, to the source. From \citet{bel97,mun99} we
expect a systematic error less than a factor of 2 arising from the
uncertainty in $N_{H}$, too small to affect our conclusions. Our
estimator from PR13 is included in the Appendix for convenience.

\section{The Energy of the Ejection} It was determined
in Punsly (2012, P12), that knowledge of the time evolution of the
spectral shape associated with a changing SSA opacity (defined from
the center of the plasmoid along a line of sight to Earth), $\tau $,
greatly enhances the accuracy of plasmoid energy, $E$, estimates
because it constrains the size. The frequency and the width of the
spectral peak provide two added pieces of information at each epoch
of observation beyond the single epoch spectral index and flux
density that is traditionally used to estimate the ejected $E$
\citep{fen99,mir94}. The slowly evolving SSA opacity of the powerful
radio flares of December 1993 was considered in the context baryon
number conservation, energy conservation, synchrotron cooling times
and X-ray luminosity in P12 to eliminate uncertainty in the energy
estimates. Namely,

\begin{enumerate}
\item The evolving $\tau$ restricts the total column depth and
plasma-filled volume, which ameliorates issues associated with
filling factor that occur in a more simplified typical minimum
energy analysis.

\item A near minimum energy condition is shown to occur when the
optically thin low frequency emission is near maximum based on the
constraint of energy conservation and the synchrotron cooling times
of a plasmoid with an evolving $\tau$ (see Section 5 and Figure 14
of P12). The peak occurs when the optical depth at 2.3 GHz,
$\tau_{2.3}$, $\approx 0.1$.

\item  Baryon number conservation and synchrotron cooling times in
combination with the evolving $\tau$ indicate that a large protonic
component requires the jet to begin nonmagnetic with most of the
energy in mechanical form (see Section 4 and Figures 12 and 13 of
P12). Yet there is insufficient radiation or thermal pressure to
initiate this outflow. Thus, only the leptonic dominated branch in
solution space can be integrated back to the source. These begin
magnetically dominated and magnetic energy is converted to
mechanical energy as equipartition is approached.

\item The magnitude of the 1.4 GHz emission requires that the energy
spectrum of electrons extend to a minimum energy, $U_{min} < 6
m_{e}c^{2}$ (see Section 5 of P12).

\end{enumerate}
In summary, the proton content is minimal, the plasmoid attains a
near minimum energy condition, $\tau_{2.3} \approx 0.1$ and
$U_{min}\approx m_{e}c^{2}$, when the optically thin flux at 2.3
GHz, $S_{\mathrm{thin}}(2.3)$, is near maximum. As in n PR13 we
invoke one assumption: the detailed modeling of the time evolution
of the radio flares from P12 can be used as a template for the time
evolution of other plasmoids with less supporting data. In this
study, we have the advantage of the complete spectral shape
including the SSA turnover. So it is possible to solve for
$\tau_{2.3}$ at the epoch of observation. Thus, this calculation is
more accurate than those of PR13 for which complete spectral data
does not exist.
\begin{figure}
\includegraphics[width= 85 mm, angle= 0]{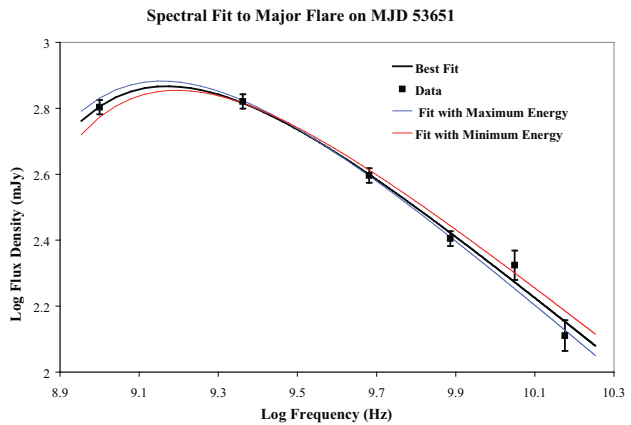}
\caption{Fits to the RATAN-600 data on MJD 53651 after core flux
subtraction. The Ryle 15 GHz data is one day after that in Figure
1.}
\end{figure}
\begin{figure}
\includegraphics[width= 85 mm, angle= 0]{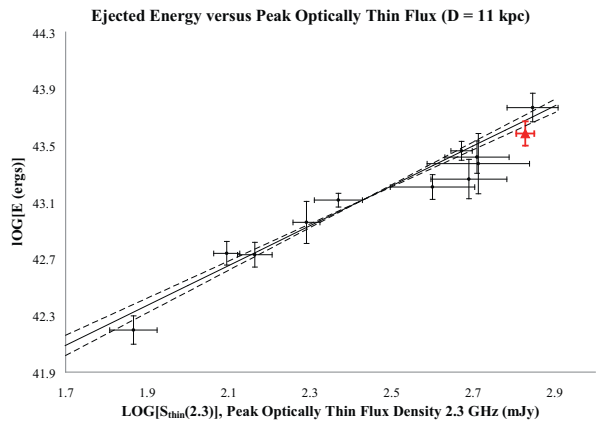}
\caption{The energy, E, estimate derived here for the 53651 radio
flare (red diamond) from the complete spectrum in Figure 2 compared
to the empirical fit from Punsly and Rodriguez (2013b) of $E$ versus
the peak $S_{\mathrm{thin}}(2.3)$, the black data points. The linear
least squares fit with uncertainty in both variables is $\log
[E(\mathrm{ergs})] = (39.69\pm 0.24) + (1.41\pm 0.09)\log
[S_{\mathrm{thin}}(2.3)/1 \mathrm{mJy}]$. The dashed lines result
from combinations of the minimum and maximum fitted values of the
coefficients as in Reed (1988) and represent the steepest and
flattest fits consistent with the data.}
\end{figure}

\par  Except when noted, a fiducial distance to GRS 1915+105 of D =11 kpc is assumed
throughout the manuscript. However, we consider every plausible
value of D and the corresponding dependent Doppler factor, $\delta$,
because of the large systematic uncertainty in $\delta$. The Doppler
factor is given in terms of $\overline{\Gamma}$, the Lorentz factor
of the outflow; $\beta$, the three velocity of the outflow and the
angle of propagation to the line of sight, $\theta$;
$\delta=1/[\overline{\Gamma}(1-\beta\cos{\theta})]$ \citep{lin85}.
In order to estimate $\delta$ from D, we assume that the kinematic
results from \citet{fen99} are common to the entire time frame from
1997 to 2005 as evidenced by interferometric observations of
multiple radio flares \cite{dha00,mil05}. The \emph{intrinsic} flux
density is $S_{\mathrm{thin}}(2.3)\delta^{-(3+\alpha)}$. As D is
varied from 10.5 kpc to near the maximum kinematically allowed value
of 11 kpc, the intrinsic spectral luminosity changes by a factor
$\approx(11/10.5)^{2}(0.33/0.56)^{4}=(1/7.5)$ which equates to a
reduction of $E$ by a factor $\sim $ 5 - 6.
\par Figure 2 shows the spectrum from RATAN-600 data provided by S. Trushkin (private
communication 2013). In addition there is a 15 GHz data point from
the Ryle archives. Before fitting the data, we want to extract the
spectrum of the ejected plasmoid from the flat spectrum core. The
method from Section 3.2 of PR13 assumes that the random variations
in the 15 GHz light curve that occur after the radio flare has risen
to its peak represent optically thick time variations of the core.
This estimation is computed by first performing a linear fit to the
15 GHz flux density from 56351.62 to 56351.91 (a day after the data
presented in Figure 1). The magnitude of the standard deviation of
the residuals from the linear fit to the data (15 mJy) is considered
to be the time averaged radio core flux density at 15 GHz (flat
spectrum: $\alpha=0$ is assumed). For such a strong radio flare this
is just a small pedantic refinement.
\par We fitted the spectral energy distribution (SED) by the methods described by Equations (1-6).
We find three fits to the data by minimizing $\chi^{2}$. The black
curve is the nominal fit to the data points. This best fit is a
powerlaw spectral luminosity with spectral index $\alpha=0.98$ that
is transferred through an SSA opacity with $\tau_{2.3}=0.33$. The
blue curve represents the fit which has the maximum plasmoid energy,
it uses the maximum value of the flux density at 1 GHz and 2.3 GHz
(top of the error bars) and the minimum value of the flux density
(bottom of the error bars) at the high frequency points. This fit
has $\alpha= 1.03 $ and $\tau_{2.3}=0.35$. The red curve represents
the fit which has the minimum plasmoid energy, it uses the minimum
value of the flux density at 1 GHz and 2.3 GHz and the maximum value
of the flux density at the high frequency points. This fit has
$\alpha= 0.94 $ and $ \tau (2.3) = 0.33$. Using the nominal value
and the maximum and minimum fits, one can compute self consistent
solutions to the Equations (1- 6) in P12 and below.

\par The relationships are expressed in observed quantities designated with a subscript, ``o".
Taking the standard result for the SSA attenuation coefficient in
the plasma rest frame and noting that the frequency obeys $\nu =
\nu_{o} / \delta$, from \citet{rey96,gin69}
\begin{eqnarray}
&& \mu(\nu)=\frac{3^{\alpha +
1}\pi^{0.5}g(p)e^{2}N_{\Gamma}}{8m_{e}c}\left(\frac{eB}{m_{e}c}\right)^{(1.5
+
\alpha)}\left(\frac{\delta}{\nu_{o}}\right)^{(2.5 + \alpha)} \\
&& g(n)= \frac{\Gamma[(3n + 22)/12]\Gamma[(3n + 2)/12]\Gamma[(n +
6)/4]}{\Gamma[(n + 8)/4]}\;.
\end{eqnarray}
This equation derives from an assumed powerlaw energy distribution
for the relativistic electrons, $ N(U)= N_{\Gamma}U^{-n}$, where the
radio spectral index $\alpha = (n-1)/2$ and $U$ is the energy of the
electrons in units of $m_{e}c^{2}$. The radiative transfer equation
was solved in \citet{gin69} to yield the following parametric form
for the observed flux density, $S_{\nu}$, from the SSA source,
\begin{eqnarray}
&& (S_{\nu})_{o} = \frac{S_{1}\nu^{-\alpha}}{R\mu(\nu)} \times
\left(1 - e^{-\mu(\nu) R}\right)\;,
\end{eqnarray}
where R is the radius of the spherical region in the rest frame of
the plasma and $S_{1}$ is a normalization factor. In the spherical,
homogeneous approximation, one can make a simple parameterization of
the SSA attenuation coefficient, $\mu(SSA)=\mu_{1}\nu_{o}^{-(2.5 +
\alpha)}$. If one assumes that the source is spherical and
homogeneous then there are three unknowns in Equation (3),
$R\mu_{1}$, $\alpha$ and $S_{1}$; $(S_{\nu})_{o}$ is determined by
observation. There is a finite range of physical parameters that are
consistent with these spectral fits. To make the connection, one
needs to relate the observed flux density in Equation (3) to the
local synchrotron emissivity within the plasma. The synchrotron
emissivity is given in \citet{tuc75},
\begin{eqnarray}
&& j_{\nu} = 1.7 \times 10^{-21} (4 \pi N_{\Gamma})a(n)B^{(1
+\alpha)}(4
\times 10^{6}/ \nu)^{\alpha}\;,\\
&& a(n)=\frac{\left(2^{\frac{n-1}{2}}\sqrt{3}\right)
\Gamma\left(\frac{3n-1}{12}\right)\Gamma\left(\frac{3n+19}{12}\right)
\Gamma\left(\frac{n+5}{4}\right)}
       {8\sqrt\pi(n+1)\Gamma\left(\frac{n+7}{4}\right)} \;.
\end{eqnarray}

\par
One can transform this to the observed flux density,
$(S_{\nu}(\nu_{o}))_{o}$, in the optically thin region of the
spectrum using the relativistic transformations from \citet{lin85},
\begin{eqnarray}
 && (S_{\nu}(\nu_{o}))_{o} = \frac{\delta^{(3 + \alpha)}}{4\pi D^{2}}\int{j_{\nu}^{'} d V{'}}\;,
\end{eqnarray}
where $j_{\nu}^{'}$ is evaluated in the plasma rest frame at the observed frequency.

\par Solving Equations (1-6) simultaneously yields an infinite number
of solutions for a given D and $\delta$ that are parameterized by
$\tau$, $R$, $N_{\Gamma}$ and $B$. Assuming that
$S_{\mathrm{thin}}(2.3) = 675 \;\mathrm{mJy}$ (Figure 2) is near the
peak $S_{\mathrm{thin}}(2.3)$ (based on its large value and the
steep spectral index at higher frequencies), according to finding
ii) above from P12, the solution with minimum energy will be close
to the physical solution. Using this insight to compute the energy
corresponding to the three fits in Figure 2 yields the (Earth frame)
energy for $D$ = 11 kpc, $\delta=0.33 $ and $\overline{\Gamma} =5.0
$ (see \citet{fen99}),
\begin{equation}
E = (3.84 \pm 0.82) \times 10^{43} \mathrm{ergs}\;.
\end{equation}
Similar expression can be found for other values of $D$ (as in
PR13). The corresponding $\delta$ and $\overline{\Gamma}$ can be
determined from the kinematics derived from interferometric
measurements per the methods of \citet{fen99}.
\par Figure 3 compares the value of $E$ in Equation (7) to that
expected from a peak $S_{\mathrm{thin}}(2.3) = 675 \;\mathrm{mJy}$
based on the best fit estimator from \citet{pun14}. The fit to the
data is by the method of list least squares with uncertainty in both
variables \cite{ree89}. The dashed lines result from combinations of
the minimum and maximum fitted values of the coefficients as in
\citet{ree89} and represent the steepest and flattest fits
consistent with the data. The radio flare on 53651 lies just below
the curve but within 1 $\sigma$ uncertainty. This is expected from
the conclusions of P12 noted in finding ii): the plasmoid should
approach minimum energy at $\tau_{2.3} \approx 0.1$, but
$\tau_{2.3}=0.33 - 0.35$ in the spectral fits. Consequently, if the
results of P12 are appropriate then the plasmoid has not quite
reached a minimum energy configuration during the observation and
the energy should be slightly elevated above minimum energy. This is
consistent with Figure 3, if the energy were slightly elevated, the
data would lie closer to the fitted curve from \citet{pun14}.

\begin{figure}
\includegraphics[width=85 mm, angle= 0]{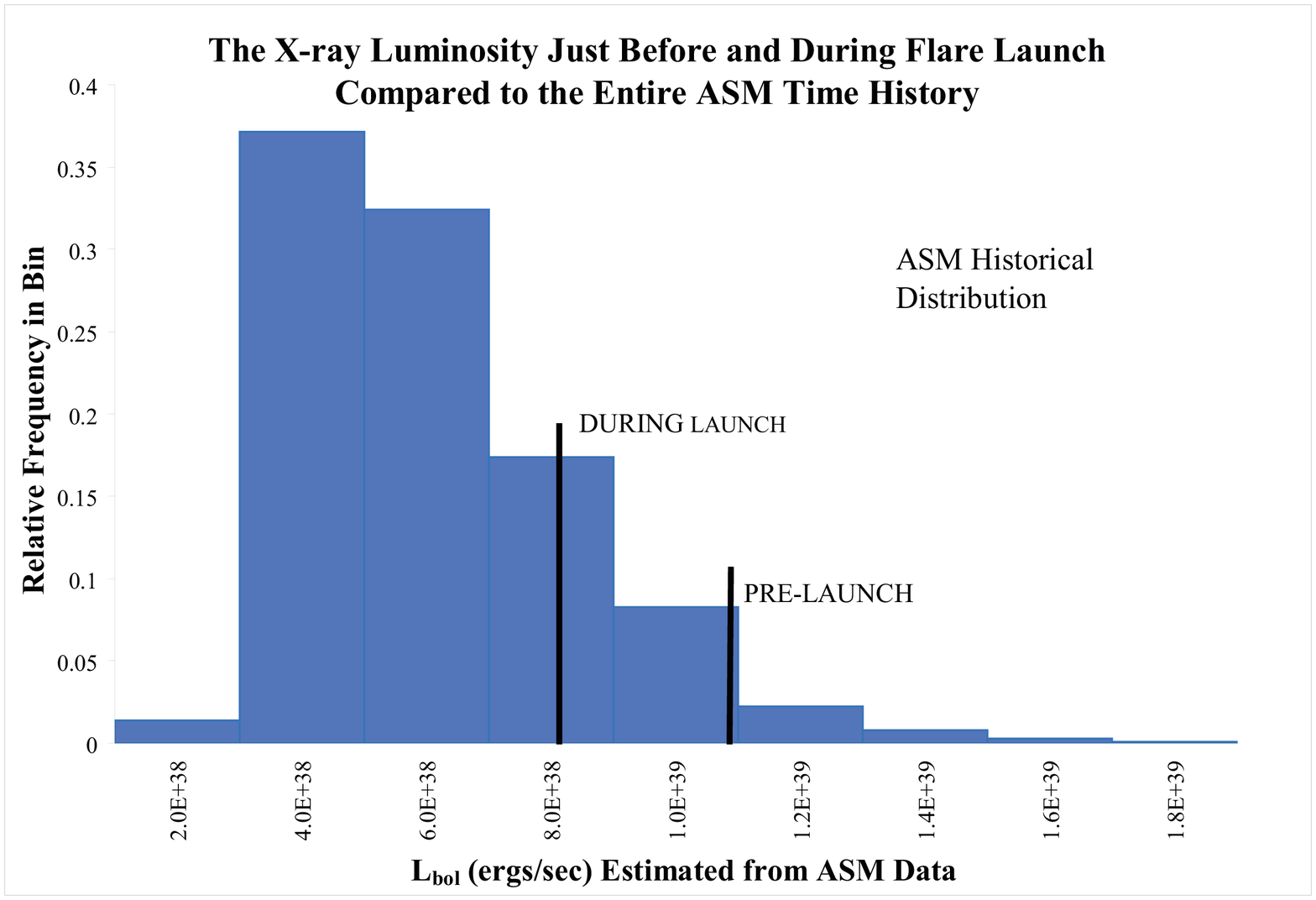}
\includegraphics[width= 85 mm, angle= 0]{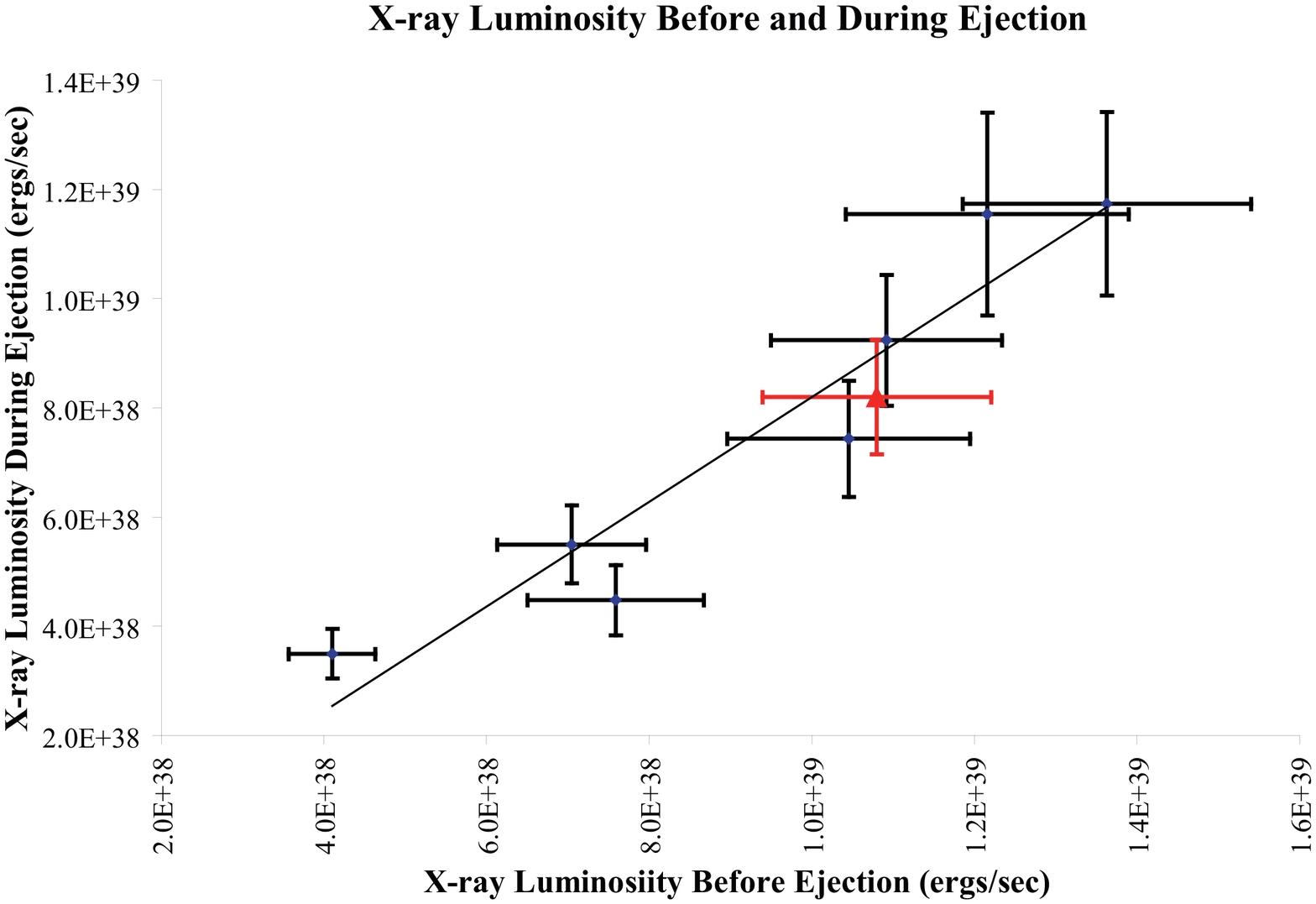}
\caption{The top frame shows a histogram of the historic
distribution of $L_{\mathrm{bol}}$ based on $>76,000$ ASM
observations. The elevated levels of $L_{\mathrm{pre-flare}}$ and
$L_{\mathrm{rise}}$ for the 53651 radio flare are indicated by the
black vertical lines. The red diamond in the bottom frame shows that
$L_{\mathrm{pre-flare}}$ and $L_{\mathrm{rise}}$ for the 53651 radio
flare obey the same correlation determined by the radio flares from
PR 13 in black. The linear fit is $L_{\mathrm{rise}}=
0.96L_{\mathrm{pre-flare}} - 1.39\times 10^{38}\mathrm{ergs/sec}$.}
\end{figure}

\section{Comparison to the Results of PR13}
The top frame in Figure 4 shows that the X-ray luminosity from
Figure 1 is elevated just (less than 4 hours) before ejection,
$L_{\mathrm{pre-flare}}$, and also so is the time average X-ray
luminosity during ejection, $L_{\mathrm{rise}}$, relative to the
historical distribution of $L_{\mathrm{bol}}$ as was shown for other
strong radio flares in PR13 (the ASM historical distribution is from
Figure 16 of PR13). The bottom frame of Figure 4 shows that the data
for the 53651 radio flare follows the trend of other radio flares
noted in Table 5 of PR13, $L_{\mathrm{pre-flare}}$ is highly
correlated with $L_{\mathrm{rise}}$.
\par Figure 5 plots the 56351 radio flare data on the background of the
correlations of E with $L_{\mathrm{pre-flare}}$ and
$L_{\mathrm{rise}}$ from PR13. Figure 5 indicates that these
potentially important physical connections between the ejection and
the accretion state gain further support from the particular case of
53651. The radio flare was not included in PR13 because there is no
15 GHz coverage to indicate the end of the plasmoid ejection
episode, so we cannot estimate the power required to launch the
plasmoid, $Q$ - a necessary condition for inclusion in PR13. Thus,
we cannot explore the correlations with $Q$ and other parameters
here.

\begin{figure}
\includegraphics[width= 85 mm, angle= 0]{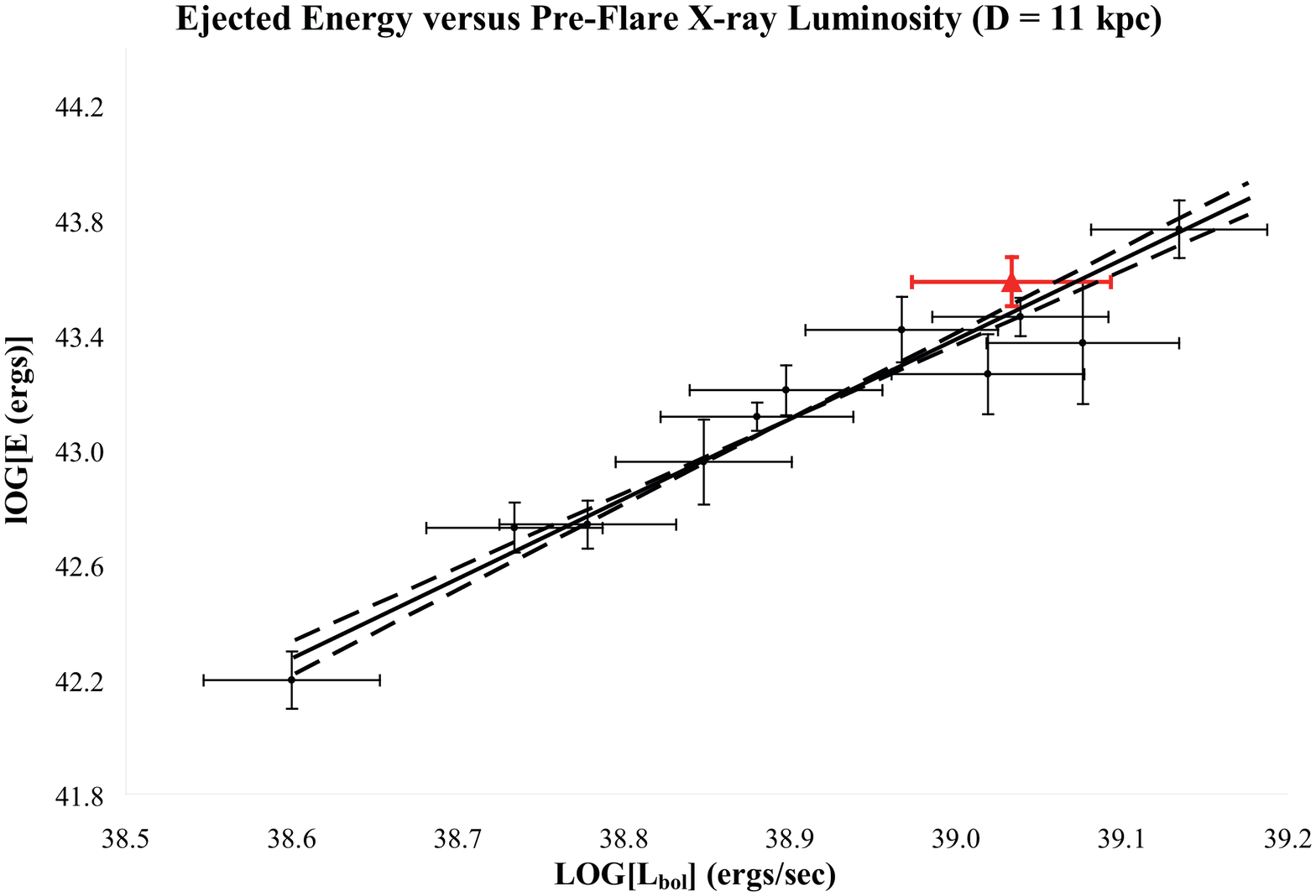}
\includegraphics[width= 85 mm, angle= 0]{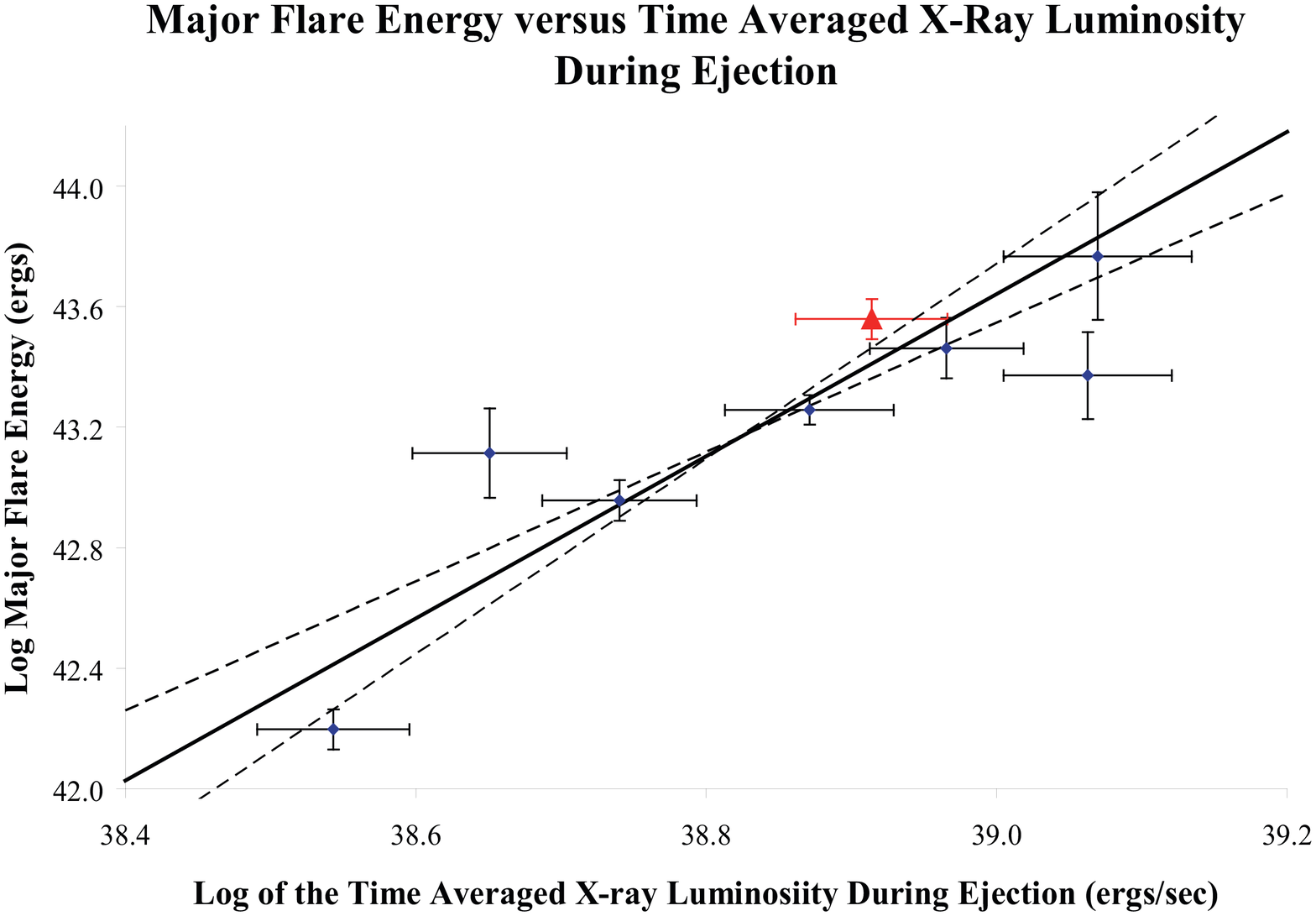}
\caption{The top (bottom) frame illustrates the correlation between
E and $L_{\mathrm{pre-flare}}$ ($L_{\mathrm{rise}}$). The data point
for 53651 is a red diamond and the fit by the method of least
squares with uncertainty in both variables is from the data in PR13
in black. The dashed lines result from combinations of the minimum
and maximum fitted values of the coefficients as in Reed (1988) and
represent the steepest and flattest fits consistent with the data.
The empirical fits are
$\log[E(\mathrm{ergs})] = -(65.07 \pm 7.76) + (2.78 \pm
0.20)\log[L_{\mathrm{pre-flare}}(\mathrm{ergs/sec})]$ and
$\log[E(\mathrm{ergs})] = -(61.31 \pm 21.23) + (2.69 \pm
0.55)\log[L_{\mathrm{rise}}(\mathrm{ergs/sec})]$. Figures 4 and 5 indicate that $L_{\mathrm{rise}}$ is a little
smaller than expected from the PR13 sample.}
\end{figure}

\section{Conclusion} In this paper, we explore the energy of the
plasmoid responsible for the major radio flare from GRS~1915+105 on
MJD 53651 and the X-ray state just before and during ejection.
Comparing to the historical distribution of flare energy in Figure 2
of \citet{pun14} indicates that this is the second most energetic
flare observed since 1996. The value of $E$ is slightly larger than
what we estimated for the large flare of April 8, 2003 from the
spectral data in \citet{fuc04}. Due to uncertainty in the estimates,
we cannot say definitively which is larger. The analysis presented
here confirms the calibration of the estimator of $E$ computed from
$S_{\mathrm{thin}}(2.3)$ in \citet{pun14} and in Figure 3 here. This
study also provides more evidence to support the strong correlations
from PR13 between E and $L_{\mathrm{pre-flare}}$ and E with
$L_{\mathrm{rise}}$ (Figure 5) as well as $L_{\mathrm{pre-flare}}$
and $L_{\mathrm{rise}}$ (Figure 4). We also find more evidence of
the finding of PR13 that $L_{\mathrm{pre-flare}}$ and
$L_{\mathrm{rise}}$ are elevated for strong flares (Figure 4). It is
important to note that the validity of the correlations is not
affected by the two main systematic uncertainties, $D$ (and the
implied changes to $\delta$ and $\overline{\Gamma}$) and $N_{H}$.
The curves in Figures 3- 5 just shift as shown in Table 5 of PR13.
\par The correlations of the accretion state preceding and during major flare ejections
found in PR13 and supported by the data presented here constrain the
physics of the mechanism that launches superluminal plasmoids. The
fact that the power of major flare ejections is correlated with an
elevated X-ray luminosity combined with the fact that the X-ray
luminosity of GRS 1915+105 is also one of the highest of any known
microquasar, \citet{don04}, leads to the obvious speculation that
GRS 1915+105 is a prolific source of major ejections as a
consequence of the high emissivity of the accretion flow. Since GRS
1915+105 radiates at a significant fraction of the Eddington
luminosity and it produces relativistic outflows, it is also natural
to consider the hypothesis that it is a small scale radio loud
quasar. In \citet{pun14}, we considered this idea in the context of
numerical simulations of accretion flows onto spinning black holes.
We could not exclude the possibility that the ejections are driven
by the accretion flow proper, but we could constrain the physics of
black hole driven jets. In particular, we found:

\begin{itemize}
\item The high luminosity of the accretion flow before and during
plasmoid launch excludes accretion states that are obstructed by an
overabundance of magnetic flux since this suppresses the source of
local dissipation, the magneto-rotational instability. In
particular, the so-called MADs (magnetically arrested accretion) and
MCAFs (magnetically choked accretion flows) that have been developed
in \citet{mck12} would not produce the observed high luminosity accretion flows.
\item If there is an analogy to radio loud quasars then the distance
to GRS 1915+105 must greater than 10.7 kpc, otherwise the time
averaged ejected power is too small.
\item If there is a black hole spin related explanation of the power source for the
major ejections then there is only one consistent set of existing
3-D numerical solutions. These are characterized by three factors.
An accretion flow that is far below the saturation point for large
scale magnetic flux. Thus, the magnetic field strength near the
black hole is governed by the pressure of the accretion flow. This
naturally correlates the power required to launch the ejection with
the accretion rate and X-ray luminosity. Secondly, large scale
magnetic flux must thread the inner regions of the accretion flow in
the ergosphere (the active region of the black hole geometry). This
results in the ergospheric disk jet that occurs in the 3-D
simulations that are described in detail in \cite{pun10}. Thirdly,
if FR II quasars are scaled up version of GRS 1915+105, the data
are consistent with numerical models when they contain an
ergospheric disk jet and the BH spin $a/M> 0.984$. This result is
intriguing because it agrees with the value of $a/M =0.99\pm 0.01$
that was estimated from a completely independent method, study based on
X-ray spectra of GRS~1915+105 \citep{mcc06,blu09}.
\end{itemize}
The need for a high bolometric luminosity prior to ejection in GRS
1915+105 and the fact that this source may be a small scale FR II
also opens interesting possibilities for the study of radio loud
AGNs. By following the UV-X-ray activity of the latter it should
then be possible to predict the onset of discrete ejection and plan
follow up radio observations more easily (due to the longer time
scale in AGN). This in turn could help us obtain more stringent
constrains on the jet properties in different types of accreting
back holes
\section*{Acknowledgments}
JR acknowledges funding support from the French Research National
Agency : CHAOS project ANR-12-BS05-0009
(http://www.chaos-project.fr)

\appendix
\section{Estimator for $L_{bol}$}
The estimator of PR13 is defined in terms of the three ASM bins,
1.2--3~keV, 3--5~keV and 5--12~keV, bin 1, bin 2 and bin 3,
respectively. The counts rates in each bin are defined
\begin{eqnarray}
&& C1 \equiv \mathrm{cts/s \; in \; bin\;1} \;, \\
&& C2 \equiv
\mathrm{max}(5.75, \mathrm{cts/s \; in \; bin\;2})\;,\\
&& C3 \equiv \mathrm{cts/s \; in \; bin\;3}\;.
\end{eqnarray}
The lower limit in the expression for C2 arises from a systematic
error that occurs in bin 2 sporadically as discussed at length in
PR13. Define the fluxes in the 3 bins
\begin{eqnarray}
&& F1= C1(3.92 \times 10^{-10})\mathrm{ergs/s-cm^2} \;, \\
&& F2= C2(3.14 \times 10^{-10})\mathrm{ergs/s-cm^2}\;,\\
&& F3= C3(4.61 \times 10^{-10})\mathrm{ergs/s-cm^2}\;,
\end{eqnarray}
We also define softness ratios
\begin{eqnarray}
&& SR1=F1/F2=0.93(C1/C2) \;, \\
&& SR2=F2/F3=0.68(C2/C3)\;.
\end{eqnarray}
Finally, we write the estimator of the intrinsic flux of
GRS~1915+105 from 1.2 - 50 keV from PR13, $F_{\mathrm{intrinsic}}$ ,
\begin{eqnarray}
&&  F_{\mathrm{intrinsic}}=0.561\left[(4.363F1(SR1)^{0.2772})\right]\nonumber \\
&&+0.561\left[(1.3767F2(SR1)^{0.0255}) +F3(1+2.730e^{-(2.114SR2)})\right] \nonumber \\
&& + 1.25 \times 10^{-8} \mathrm{ergs/s}/\mathrm{cm}^{2}\;,\;
\mathrm{if} \; SR1<1\;,
\end{eqnarray}
\begin{eqnarray}
&& F_{\mathrm{intrinsic}} = 0.478\left[(4.363F1(SR1)^{0.2772})\right]\nonumber\\
&& +0.478 \left[(1.3767F2(SR1)^{0.0255}) +F3(1+2.730e^{-(2.114SR2)})\right]\nonumber \\
&& + 1.08 \times 10^{-8} \mathrm{ergs/s}/\mathrm{cm}^{2}\;,\;
\mathrm{if} \; SR1>1\;.
\end{eqnarray}

\begin{thebibliography}{99}
\bibitem[\protect\citeauthoryear{Belloni et al}{1997}]{bel97}Belloni, T. et al 1997, ApJL 488 109
\bibitem[\protect\citeauthoryear{Belloni et al}{2000}]{bel00}Belloni, T., Klein-Wolt, M., Mendez, M., van der Klis, M., van Paradijs, J. 2000, A \& A 355
271
\bibitem[\protect\citeauthoryear{Blum et al}{2009}]{blu09}Blum, J. et al 2009, ApJ 706 60.
\bibitem[\protect\citeauthoryear{Corbel et al}{2000}]{cor00}Corbel, S. et al 2000, A \& A 359 251
\bibitem[\protect\citeauthoryear{Corbel et al}{2003}]{cor03}Corbel, S., Nowak, M. A., Fender, R. P., Tzioumis, A. K., Markoff, S. 2003, A \& A 400 1007
\bibitem[\protect\citeauthoryear{Dhawan et al}{2000}]{dha00}Dhawan, V., Mirabel, I.F., Rodriguez, L. 2000, ApJ 343 373
\bibitem[\protect\citeauthoryear{Done et al}{2004}]{don04}Done, C., Wardzinski, G., Gierlinski, M. 2004, MNRAS 349 393
\bibitem[\protect\citeauthoryear{Fender et al}{1999}]{fen99}Fender, R. et al, 1999, MNRAS 304 865
\bibitem[\protect\citeauthoryear{Fuchs et al}{2004}]{fuc04}Fuchs, Y. et al, 2004, Proc. of the 5th INTEGRAL Workshop (ESA) http://xxx.lanl.gov/abs/astro-ph/0404030
\bibitem[\protect\citeauthoryear{Gallo et al}{2003}]{gal03}Gallo, E., Fender, R., Pooley, G. 2003, MNRAS 344 60
\bibitem[\protect\citeauthoryear{Ginzburg \& Syrovatskii}{1969}]{gin69} Ginzburg, V. and Syrovatskii, S. 1969,
  Annu. Rev. Astron. Astrophys. \textbf{7} 375
\bibitem[\protect\citeauthoryear{Klein-Wolt et al.}{2001}]{kle02}Klein-Wolt et al., 2002, MNRAS 331 745
\bibitem[\protect\citeauthoryear{Lind \& Blandford}{1985}]{lin85}Lind, K., Blandford, R.
1985, ApJ \textbf{295} 358
\bibitem[\protect\citeauthoryear{McClintock et al}{2006}]{mcc06}McClintock, J. et al 2006, ApJ 652 518.
\bibitem[\protect\citeauthoryear{Mckinney et al}{2012}]{mck12}McKinney, J., Tchekhovskoy, A., Blandford, R. 2012, MNRAS 423
3083
\bibitem[\protect\citeauthoryear{Miller-Jones et al}{2005}]{mil05}Miller-Jones, J. et al 2005, MNRAS 363 867
\bibitem[\protect\citeauthoryear{Mirabel \& Rodriguez}{1994}]{mir94}Mirabel, I.F., Rodriguez, L. 1994, Nature 371 46
\bibitem[\protect\citeauthoryear{Mirabel et al}{1998}]{mir98}Mirabel, I.F. et al 1998, A \& A 330 L9
\bibitem[\protect\citeauthoryear{Muno et al}{1999}]{mun99}Muno, M., Morgan, E, Remillard, R.  1999, ApJ 527 321
\bibitem[\protect\citeauthoryear{Punsly}{2012}]{pun12} (P12) Punsly, B. 2012, ApJ 746 91
\bibitem[\protect\citeauthoryear{Punsly et al}{2010}]{pun10} Punsly, B., Igumenshchev, I. V., Hirose, S. 2010, ApJ 704, 1065
\bibitem[\protect\citeauthoryear{Punsly \& Rodriguez}{2013a}]{pun13} (PR13) Punsly, B., Rodriguez J. 2013a, ApJ 764 173
\bibitem[\protect\citeauthoryear{Punsly \& Rodriguez}{2013b}]{pun14} Punsly, B., Rodriguez J. 2013b,
ApJ 770 99
\bibitem[\protect\citeauthoryear{Reed}{1989}]{ree89} Reed, B. 1989, Am. J. Phys. 57 642
\bibitem[\protect\citeauthoryear{Reynolds et al.} {1996}]{rey96}Reynolds, C. S., Fabian, A., Celloti, A., Rees, M.
1996, MNRAS \textbf{283} 873
\bibitem[\protect\citeauthoryear{Rodriguez, J., Hannikainen, D., Shaw, S., et al.} {2008a}]{rod08} Rodriguez, J., Hannikainen, D., Shaw, S., et al.  2008, ApJ 675
1436
\bibitem[\protect\citeauthoryear{Rodriguez, J.,  Shaw, S., Hannikainen, D., et al.} {2008b}]{rod09} Rodriguez, J.,  Shaw, S., Hannikainen, D., et al. 2008, ApJ 675
1449
\bibitem[\protect\citeauthoryear{Rushton et al.} {2010}]{rus11}Rushton, A., Spencer, E., Fender, R. and Pooley, G. 2010, A\& A 524 29
1449
\bibitem[\protect\citeauthoryear{Tucker}{1975}]{tuc75}Tucker, W. 1975, \emph{Radiation Processes in
    Astrophysics} (MIT Press, Cambridge).
\end{thebibliography}
\end{document}